\documentclass[twocolumn]{revtex4}

\usepackage{amsmath}
\usepackage{amsthm}
\usepackage{amssymb}
\usepackage{braket}
\usepackage{graphicx}
\newcommand{\tth}{th }

\newcommand{\mt}{\mathrm}
\newcommand{\ud}{\mathrm{d}}
\newcommand{\vv}[1]{\ensuremath{\mathbf{#1}}}
\newcommand{\NN}{\ensuremath{\mathbb{N}}}
\newcommand{\lk}{\left(}
\newcommand{\rk}{\right)}

\newcommand{\pure}[1]{\ensuremath{\ket{#1}\bra{#1}}}

\newcommand{\signal}{\ensuremath{^{(\mathrm{s})}}}
\newcommand{\decoy}{\ensuremath{^{(\mathrm{d})}}}
\newcommand{\sigdec}{\ensuremath{^{(\mathrm{s/d})}}}
\newcommand{\lb}{\mathrm{LB}}
\newcommand{\LB}{\mathrm{LB}}
\newcommand{\ub}{\mathrm{ub}}
\newcommand{\olb}{\ensuremath{_{1,\LB}}}
\newcommand{\mub}{\ensuremath{_{\mathrm{M,UB}}}}
\newcommand{\owc}{\ensuremath{_{1,\mathrm{lb}}}}
\newcommand{\mwc}{\ensuremath{_{\mathrm{M,ub}}}}
\newcommand{\hnix}{\ensuremath{^{}}}
\newcommand{\multi}{\ensuremath{_{\mathrm{M}}}}

\newcommand{\sig}{\ensuremath{_{\mathrm{s}}}}
\newcommand{\id}{\ensuremath{_{\mathrm{i}}}}

\DeclareMathOperator*{\argmax}{argmax}

\DeclareMathOperator{\sech}{sech}

\def\clap#1{\hbox to 0pt{\hss#1\hss}}
\def\mathclap{\mathpalette\mathclapinternal}
\def\mathclapinternal#1#2{%
  \clap{$\mathsurround=0pt#1{#2}$}%
}

\begin{document}
  
\title{Multi-mode states in decoy-based quantum key distribution protocols}

\author{Wolfram \surname{Helwig}}
\email[]{wolfram.helwig@physik.uni-erlangen.de} 
\affiliation{Max Planck Institute for the Science of Light}
 
\author{Wolfgang \surname{Mauerer}}
\affiliation{Max Planck Institute for the Science of Light}
  
\author{Christine \surname{Silberhorn}}
\affiliation{Max Planck Institute for the Science of Light}
  
\begin{abstract}
Every security analysis of quantum key distribution (QKD) relies on a
faithful modeling of the employed quantum states.  Many photon sources, like
for instance a parametric down conversion (PDC) source, require a multi-mode
description, but are usually only considered in a single-mode
representation. In general, the important claim in decoy-based QKD protocols
for indistinguishability between signal and decoy states does not hold for
all sources. We derive new bounds on the single photon transmission
probability and error rate for multi-mode states, and apply these bounds to
the output state of a PDC source. We observe two opposing effects on the
secure key rate. First, the multi-mode structure of the state gives rise to
a new attack that decreases the key rate. Second, more contributing modes
change the photon number distribution from a thermal towards a Poissonian
distribution, which increases the key rate.
\end{abstract}
\date{\today}

\maketitle

\section{Introduction}
The security of classical cryptography is based on the high computational
complexity
of the decryption process combined with the
condition that the adversary only has a limited amount of computational power.
In contrast, quantum key distribution (QKD) allows two parties, Alice and Bob,
to share a secret key that is inaccessible to an eavesdropper Eve
whose power is only limited by the laws of quantum physics. In 1984, Bennett
and Brassard introduced the first QKD protocol BB84 \cite{Bennett1984}. It is
still the most commonly used protocol, although many more have been proposed
since then \cite{Gisin2002, Duvsek2006, Mauerer2008a}.

This first theoretical proposal assumed perfect
devices, namely single-photon sources and error-free transmission and
detection. With the development of sophisticated security proofs, these
restrictions could gradually be lessened. First, security has been proved in
the presence of noise \cite{Mayers1998, Shor2000}. In the next step, the
necessity of single-photon sources has been taken out of the equation
\cite{Gottesman2004}. This, however, reduced the achievable key rate
drastically, because multiphoton events give rise to the photon number splitting
(PNS) attack \cite{Lutkenhaus2000, Brassard2000, Lutkenhaus2002}, which Alice
and Bob cannot distinguish from natural losses.

This issue can be resolved with the decoy method, which was introduced by
Hwang \cite{Hwang2003}, and has been further developed to a practically
realizable form by several researchers \cite{Lo2005, Wang2005a, Ma2005}.
In this method, additional \emph{decoy} states with a different photon number
distribution than the primary signal states are randomly introduced. It is
crucial that decoy states share all other physical characteristics of the
signal so that Eve cannot distinguish between decoy and signal. Consequently,
the decoys are affected by the PNS attack in the same way as the signal states,
and this perturbation of the system reveals Eve's presence.
Since Eve has to design her attack in a way that cannot be detected by
Alice and Bob (otherwise the protocol is aborted), her attack possibilities are
drastically limited when she is confronted with a decoy protocol. This enables
Alice and Bob to achieve an improved key rate.

The important assumption that Eve cannot
distinguish between photons arising from signal and decoy states is trivially
fulfilled  for a single-mode description where
all photons are created by the same creation operator. However, this model does
not match experimental reality well.
Hence, in this paper, we treat the scenario where
photons are excited into many different modes and the excitation probability
for each mode differs between signal and decoy states.
This multi-mode description is, for
instance, necessary for a realistic representation of the states created by a
parametric down conversion (PDC) source, in which case the different modes
correspond to different spectral modes \cite{Mauerer2008}.

This paper is organized as follows. 
In Sec.~\ref{section:decoy}, we review
the decoy method and introduce the notation necessary for the subsequent
analysis. Sec.~\ref{section:spectral} presents the description
of a multi-mode state with special emphasis on spectral modes for the
description of a PDC state. In Sec.~\ref{section:Y1} and \ref{section:e1},
we describe new attack possibilities when multi-mode states are used and derive
new bounds that allow us to calculate the achievable
key rate in this scenario. Sec.~\ref{section:simulation} finally applies
the analysis to the multimode PDC state and gives new bounds on the achievable
key rate.

\section{Decoy Method}
\label{section:decoy}
The security of BB84 is based on the no-cloning theorem \cite{Wootters1982},
which prevents Eve from making a copy of a transmitted single photon. 
However, the security argument is not applicable to multiphoton events, because
for these events Alice implicitly encodes the same information on all photons in
the
pulse. This, in turn, allows Eve to obtain an identical copy of Bob's state by
splitting away one of the photons. Hence only detection events arising from
single photons give a positive contribution to the secure key rate. With
current technology, Alice is not able to determine the number of photons her
source emitted. Thus Alice and Bob cannot simply ignore multiphoton events. In
this scenario, a lower bound on the secure key rate $S$ is given by
\cite{Gottesman2004, Lo2005}
\begin{equation}
  \label{eq:keyrate}
  S \geq q \left[ Y_{1}\signal p_1\signal \lk 1 - H(e_{1}\signal) \rk -
      Q\signal f(E\signal) H(E\signal) \right].
\end{equation}
Here $p_n\signal$ ($n\in \NN_0$) denotes the photon number distribution of
Alice's source,
$f(E\signal)$ is the error correction efficiency, $H(x) = -x
\log(x) - (1-x) \log(1-x)$ is the binary Shannon entropy and $q$ accounts for
incompatible basis choices of Alice and Bob. In the standard
BB84 protocol, $q = 1/2$. The overall detection probability is given by the
\emph{gain}
\begin{equation}
  \label{eq:Qsignal}
  Q\signal = Y_0\hnix p_0\signal + Y_1\signal p_1\signal + 
      Y\multi\signal p\multi\signal,
\end{equation}
where the \emph{yields} $Y_0\signal$, $Y_1\signal$ and
$Y\multi\signal$ are the detector click probabilities conditioned on emitted
zero-, one-, and multiphoton events of Alice's source, respectively.
Analogously, the zero-, one- and multi-photon error rates, $e_0$, $e_1$ and
$e\multi$, are defined as the error rates conditioned on emitted zero-, one-,
and multiphoton events, respectively. The relation to the total quantum bit
error rate (QBER) $E\signal$ is given by
\begin{equation}
  \label{eq:QBERsignal}
  Q\signal E\signal = 
    e_0\hnix Y_0\hnix p_0\signal + e_1\signal Y_1\signal p_1\signal + 
      e\multi\signal Y\multi\signal p\multi\signal.
\end{equation}
The QBER and the gain $Q\signal$ are directly accessible
from the recorded
data of the QKD protocol. However, since Alice and Bob do not know when a single
photon was sent, they cannot determine the
exact values of $Y_1\signal$ and $e_1\signal$, but need to estimate them using
worst-case assumptions. Prior to the decoy method, they had to assume
that all multiphoton events produce a click at
Bob's detector. This corresponds to $Y\multi\signal = 1$, and a lower bound on
$Y_1\signal$ can be calculated with Equation~\eqref{eq:Qsignal}. This estimate,
however, lies well below the single photon transmission probability caused by
natural losses. In addition, Alice and Bob have to assume that
all errors arise from single photon events, resulting in a very high estimate
of
$e_1\signal$. These values are actually achieved if Eve performs a photon number
splitting (PNS) attack \cite{Lutkenhaus2000, Brassard2000, Lutkenhaus2002}, and
lead to a drastically reduced key rate
\cite{Gottesman2004}.

The decoy method enables Alice and Bob to attain better estimates of
$Y_1\signal$
and $e_1\signal$. This is achieved by randomly introducing ``decoy'' states with
independent photon number distributions. For each photon number
distribution, the gain and QBER can be determined individually, resulting in
better bounds on $Y_1\signal$ and $e_1\signal$.

In this paper, we base our analysis on the so-called \emph{vacuum+weak decoy
method} \cite{Wang2005a, Ma2005}, which uses two decoy states.
Deliberately interspersing the signal stream with vacuum states ($p_n =
\delta_{n0}$, gain $Q=Y_0$) allows Alice and
Bob to determine the dark count probability $Y_0$. The second decoy state is 
of low
intensity and features a photon number distribution $p_n\decoy$ that differs
from the photon number distribution $p_n\signal$ of the regular signal.
In the following, we will
refer to this state as the decoy state, and to the state with photon number
distribution $p_n\signal$ as the signal state.
The gain for the decoy state is given by 
\begin{equation}
  \label{eq:Qdecoy}
  Q\decoy = Y_0\hnix p_0\decoy + Y_1\decoy p_1\decoy + 
      Y\multi\decoy p\multi\decoy,
\end{equation}
with the yields defined equivalently to the signal yields. The yield $Y_0$ for
an emitted zero-photon state has to be the same for all states because Eve
cannot distinguish between vacua arising from different states.
However, there is no \textit{a priori} reason for the single- and
multiphoton yields to be the same for signal and decoy. In the analyses up to
now,  it was assumed
that Eve cannot distinguish between $n$-photon events arising from signal and
decoy state, resulting in $Y_n\signal = Y_n\decoy$ ($n \in \mathbb{N}_0$).
Note that this is the assumption we will loosen in Section~\ref{section:Y1}, as 
it is generally not justified in a multi-mode description of the states. 
However, proceeding with $Y_n\signal = Y_n\decoy$, from Equations
\eqref{eq:Qsignal} and \eqref{eq:Qdecoy} the lower bound
\begin{multline}
  \label{eq:Y1sm}
  Y_{1, \lb} = 
    \frac{p_2\signal}{p_2\signal p_1\decoy - p_1\signal p_2\decoy}
      \times\\
    \Bigg(
      Q\decoy
      - \frac{p_2\decoy}{p_2\signal} Q\signal
      - \frac{p_2\signal p_0\decoy - p_0\signal p_2\decoy}{p_2\signal}
Y_0
    \Bigg)
\end{multline}
on the signal single photon yield can be derived if the
additional condition 
\begin{equation}
  \label{eq:Yrelsm}
  \frac{Y\multi\decoy}{Y\multi\signal} \leq \frac{p_2\decoy
    / p\multi\decoy}{p_2\signal / p\multi\signal}
\end{equation}
is satisfied. This is, for instance, fulfilled if both signal and decoy have a
Poissonian or thermal distribution with a lower mean photon number for the
decoy distribution.

With a similar Equation to \eqref{eq:QBERsignal} for the decoy QBER (i.e., 
$(.)\signal \rightarrow (.)\decoy$), an upper bound
on the single photon error rate of the decoy state can be calculated as
\begin{equation}
  \label{eq:e1ub}
  e_1\decoy \leq e_{1,\ub}\decoy =
    \frac{1}{p_1\decoy Y_{1,\lb}\decoy} \left[
    E\decoy Q\decoy - e_0 Y_0 p_0\decoy
    \right].
\end{equation}
The postulated indistinguishability of $n$-photon states for signal and
decoy gives $e_{1,\ub}\signal = e_{1,\ub}\decoy$ and $Y_{1,\lb}\decoy =
Y_{1,\lb}\signal$, because $Y_1\decoy = Y_1\signal$, and $e_0 = 1/2$, since a
dark count gives the wrong result 50\% of the time.

The derived bounds, Equations \eqref{eq:Y1sm} and \eqref{eq:e1ub}, are much
tighter than the worst case assumptions without decoy states. Therefore using
them in Equation~\eqref{eq:keyrate} results in a significant improvement in the
achievable secure key rate \cite{Lo2005}.

\section{The Multi-Mode State}
\label{section:spectral}
QKD analyses generally assume that Alice's output states are accurately
represented by a single-mode description as
\begin{equation}
  \label{eq:statesinglemode}
  \rho\sigdec = \sum_n p\sigdec(n) \pure{n},
\end{equation}
where $p\sigdec(n)$ denotes the signal and decoy photon number distributions.

This form intrinsically implies that all emitted photons have identical
properties, in particular they are all excited into the same spectral mode.
This appears as an appropriate
modeling for weak coherent pulses emitted by a laser as multi-mode effects can
be expected to be less critical.
However, designing a PDC source with single-mode emission
is a complex task, as it requires two output beams with
independent spatio-spectral mode structures \cite{Mosley2008}.

A type-II PDC process emits photons in two different polarization modes, called
signal and idler. The photon numbers in signal and idler modes
are strictly correlated. If PDC sources are used in "prepare and measure" QKD
protocols, only the photons of the signal mode are employed for information
encoding. 
Describing the PDC output state in a single-mode description as
$\ket{\Psi\sigdec} = \sum_n c\sigdec(n) \ket{n}\sig \ket{n}\id$, we can obtain
signal states of the form of Equation \eqref{eq:statesinglemode} by tracing
over the idler mode.

A more realistic model of the PDC output state has to account for the
multi-mode structure as follows \cite{Mauerer2008}:
\begin{equation}
  \label{eq:mmpdcstate}
  \ket{\Psi\sigdec} = \bigotimes_{k=1}^N \sum_n c\sigdec(k,n) 
    \ket{n,\xi_k}\sig \ket{n,\zeta_k}\id
\end{equation}
with
\begin{gather}
\label{eq:mmnsig}
  \ket{n, \xi_k}\sig = \frac{1}{\sqrt{n!}} \lk A_{\mt{s},\xi_k}^\dag\rk^n
    \ket{0},\\
  A_{\mt{s},\xi_k}^\dag \equiv \int \ud \omega \xi_k(\omega) a^\dag\sig(\omega).
\end{gather}
Equation~\eqref{eq:mmnsig} describes a state with $n$ photons in the
spectral mode $\xi_k(\omega)$. Here $\xi_k(\omega)$ is a set of orthonormal
functions, therefore the orthonormality condition
$\braket{n,\xi_i|m,\xi_j} = \delta_{nm} \delta_{ij}$ holds. For a detailed
description of this notation, see \cite{Rohde2007}.
Note that we consider only one spatial mode in
Equation~\eqref{eq:mmpdcstate}, which can
be achieved by using a single mode fiber. The following analysis, however,
does not depend on the mode type and thus can also be applied to states with
more than one spatial mode.

Tracing out the idler mode gives the state
\begin{equation}
  \label{eq:rhosigdec}
  \rho\sigdec = \bigotimes_{k=1}^N \sum_n p\sigdec(k,n)
    \pure{n; \xi_k}
\end{equation}
with $p\sigdec(k,n) = |c\sigdec(k,n)|^2 = \sech^2 r_k\sigdec \tanh^{2n}
r_k\sigdec$ in the
signal arm \cite{Mauerer2008}. Here the squeezing parameter $r_k\sigdec \propto
\sqrt{I\sigdec}$ is proportional to the square root of the pump intensity.
Hence, if signal and decoy states are created by pumping the crystal with
different intensities, the mode distributions for signal and decoy states will
be different and our multi-mode treatment becomes essential.

In our numerical simulations, Section~\ref{section:simulation}, we will return
to investigating the PDC state to illustrate the multi-mode effects in
decoy-based QKD systems. Note, however, that the following analysis applies
to all states of the form~\eqref{eq:rhosigdec}
and is by no means restricted to a PDC source or spectral modes.

\section{New Bound on $Y_1\signal$}
\label{section:Y1}
Recall that in a single mode description, the yields for $n$-photon signal and
decoy states are identical and thus a lower bound on $Y_1\signal$ can be
computed by Equation~\eqref{eq:Y1sm} from the known gains and photon number
distributions. 

In this section, we develop a means of computing a lower bound on $Y_1\signal$
for multi-mode states as defined in Equation~\eqref{eq:rhosigdec}. For states of
this type, $Y_n\signal = Y_n\decoy$ is no longer valid
for $n \geq 1$. Note, however, that $Y_0$ is still the same
for signal and decoys, because for zero emitted photons the
resulting state is always described by the same vacuum state and thus Eve cannot
treat these pulses differently for signal and decoys.
The derivation of the new lower bound on $Y_1\signal$ proceeds in three
steps. First, we derive a lower bound on the signal single-photon yield
$Y_1\signal$ for a given decoy single-photon yield $Y_1\decoy$. Then, we 
determine an upper bound on the decoy multiphoton yield $Y\multi\decoy$ for a
given signal multiphoton yield $Y\multi\signal$. Finally, with these two
relations, we are able to calculate a lower bound on $Y_1\signal$ for
given signal and decoy gains, $Q\signal$ and $Q\decoy$.

\textbf{Step 1 - Lower bound on $Y_1\signal$ for a given $Y_1\decoy$:}
Assume Eve has to let a certain fraction $Y_1\decoy$ of the decoy
single-photon events pass to achieve the desired decoy gain. In this step, we
are seeking the lowest possible value for the signal single-photon yield
$Y_1\signal$ that is compatible with the given $Y_1\decoy$.
Using Equation~\eqref{eq:rhosigdec}, we find the conditioned one-photon state
to be
\begin{equation}
  \label{eq:rho1sigdec}
  \rho_1\sigdec =
    \sum_k m_k\sigdec \pure{1;\xi_k},
\end{equation}
where we define the mode occupation probabilities by
\begin{equation}
  \label{eq:mksigdec}
  m_k\sigdec = \frac{p\sigdec(k,1) \prod_{i \neq k}
        p\sigdec(i,0)}{{P_1\sigdec}},
\end{equation}
and the single-photon probability is given by
\begin{equation}
  \label{eq:P1sigdec}
  P_1\sigdec = \sum_k p\sigdec(k,1) \prod_{i\neq k} p\sigdec(i,0).
\end{equation}
We have to assume that Alice does not have the technology to determine in which
mode a photon resides. Remember that she cannot even determine the total number
of
emitted photons for a given event. However, an apparatus that measures the
number of photons in each mode individually is possible in principle, because
all modes are
orthogonal. Therefore, if we want to claim unconditional security, we have to
give Eve knowledge about how many photons each mode contains. For the
single-photon case, this means that she knows in which mode the photon is.
This, in turn, allows Eve to reach different single-photon yields for signals
and decoys by selectively blocking modes if the
mode occupation probabilities $m_k$ differ for signal and decoy state.

\begin{figure}
  \centering
  \includegraphics[width=0.45\textwidth]{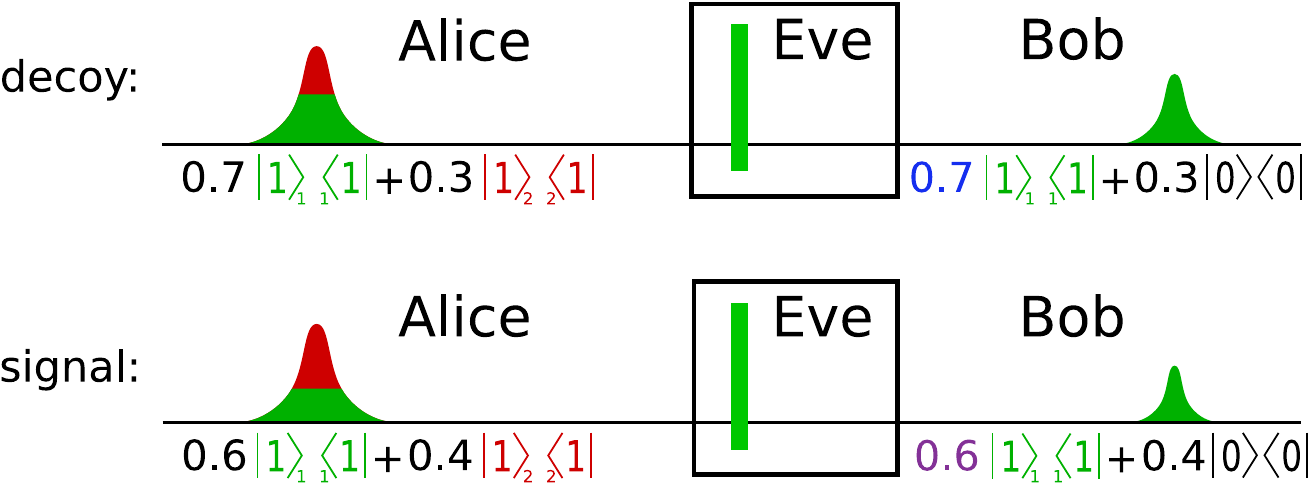}
   \caption{(Color online) Eve blocks all photons in the second mode. This
results in different
    single-photon yields for signal and decoy state.}
  \label{fig:blocksignalmode}
\end{figure}

Figure~\ref{fig:blocksignalmode} illustrates how this can be accomplished for
the case of two different modes with $m_1\decoy = 0.7$, $m_2\decoy =
0.3$, $m_1\signal = 0.6$ and $m_2\signal = 0.4$. If Eve blocks all photons in
the second mode, 70\% of the decoy single-photons event are transmitted, but
only 60\% of the signal single-photons pass through the channel. This results in
the yields $Y_1\signal = 0.6$ and $Y_1\decoy = 0.7$.

For a given decoy single-photon yield $Y_1\decoy$, we denote the smallest
possible value Eve can achieve for $Y_1\signal$ as
 $Y\owc\signal\big( Y_1\decoy\big)$.
In our example above, we have $Y\owc\signal = 6/7 \cdot Y_1\decoy$ for
$Y_1\decoy
\leq 0.7$, because in this case the photons of the first mode are sufficient to
reach $Y_1\decoy$. Thus Eve can completely block the second mode and let only
photons of the first mode pass. For $Y_1\decoy > 0.7$, Eve additionally
has to let a fraction $(Y_1\decoy - 0.7)/0.3$ of the photons in the second mode
pass to reach $Y_1\decoy$, resulting in $Y\owc\signal = 0.6 +
(Y_1\decoy - 0.7)/0.3 \cdot 0.4$.

This concept is easily extended to all $N$ modes that
contribute to the state of Equation~\eqref{eq:rho1sigdec}. Without
loss of generality, we take
$m_1\decoy/m_1\signal \geq m_2\decoy/m_2\signal \geq \ldots \geq
m_N\decoy/m_N\signal$. Only letting photons from the first mode pass results in
the smallest ratio between $Y_1\signal$ and $Y_1\decoy$, but this is only
possible if $Y_1\decoy \leq m_1\decoy$. Otherwise, additional photons from other
modes are needed to achieve the desired $Y_1\decoy$. The number $K$ of required
modes is implicitly defined by
\begin{equation}
  \sum_{k=1}^{K-1} m_k\decoy \leq Y_1\decoy < \sum_{k=1}^K m_k\decoy,
\end{equation}
such that both inequalities hold.
Eve can achieve the lowest possible value for $Y_1\signal$ by letting all
photons of the first $K-1$ modes, and a fraction $\lk Y_1\decoy -
\sum_{k=1}^{K-1} m_k\decoy\rk/m_K\decoy$ of the photons in the
$K$\tth mode pass.
This gives the desired value for the decoy single-photon yield
\begin{equation}
  Y_1\decoy = \sum_{k=1}^{K-1} m_k\decoy
      + \frac{Y_1\decoy - \sum_{k=1}^{K-1} m_k\decoy}{m_K\decoy} m_K\decoy,
\end{equation}
and the lower bound
\begin{equation}
  \label{eq:Yowc}
  Y\owc\signal\lk Y_1\decoy\rk = \sum_{k=1}^{K-1} m_k\signal
      + \frac{Y_1\decoy - \sum_{k=1}^{K-1} m_k\decoy}{m_K\decoy} m_K\signal
\end{equation}
for the signal single-photon yield for a given $Y_1\decoy$.

\textbf{Step 2 - Upper bound on $Y\multi\decoy$ for a given $Y\multi\signal$:}
In this step, we want to find the highest possible decoy multiphoton yield
that is compatible with a given signal multiphoton yield. With minor
modifications, this works out analogously to step 1, the only difference being
that we have to keep track of all possible distributions of the photons among
the modes. For this purpose, we introduce the set $Q = \{ \vv l\in \NN_0^N |
\sum_{k=1}^N l_k \geq 2 \}$.
Each member of this set represents a multiphoton event with $l_i$ photons in
the $i$\tth mode. 

Similarly to $m_k$ for the single-photon case, we define $h_{\vv l}$ as the
probability that a multiphoton event possesses the photon distribution
specified by $\vv l \in Q$. It is given by 
\begin{equation}
  \label{eq:hl}
    h_{\vv l}\sigdec =
    \frac{1}{P\multi\sigdec}
    \prod_{k=1}^N p\sigdec(k, l_k)
\end{equation}
with the multiphoton probability
\begin{equation}
  \label{eq:Pmulti}
    P\multi\sigdec = \sum_{n\geq 2} P_2\sigdec = 1 - P_0\sigdec - P_1\sigdec,
\end{equation}
where $P_n\sigdec$ denotes the convoluted photon number distribution of all
modes. Accordingly, $P_0\sigdec = \prod_{k=1}^N p\sigdec(k,0)$ and $P_1\sigdec$
is given by Equation \eqref{eq:P1sigdec}. Employing the mode distribution
probabilities of Equation~\eqref{eq:hl}, the states of
Equation~\eqref{eq:rhosigdec} conditioned on a
multiphoton event can be written as
\begin{equation}
    \rho\multi\sigdec =
      \sum_{\vv l \in Q} h_{\vv l}\sigdec
      \bigotimes_{k=1}^N \pure{l_k; \xi_k}.
\end{equation}
Again, Eve is not only allowed to make a photon number measurement but can also
determine the mode distribution $\vv l$ of a multiphoton event. Thus she can
selectively block multiphoton events with certain mode distributions. The
highest possible $Y\multi\decoy$ for a given $Y\multi\signal$ is achieved if
Eve lets only events with the highest ratio between $h_{\vv l}\decoy$ and
$h_{\vv l}\signal$ pass. To sort the mode distributions accordingly,
we define $L_1 = \argmax_{L \in Q} h_L\decoy / h_L\signal$ and recursively
\begin{equation}
L_i = \argmax_{L \in Q \setminus \{L_1, \ldots ,L_{i-1}\} }
\frac{h_L\decoy}{h_L\signal}
 \qquad \text{for } i\geq2.
\end{equation}
With that definition, we can apply the same method as in step 1. For a given
$Y\multi\signal$, we define $K$ implicitly by
\begin{equation}
  \sum_{i=1}^{K-1} h_{L_i}\signal \leq Y\multi\signal <
    \sum_{i=1}^K h_{L_i}\signal.
\end{equation}
The highest possible $Y\multi\decoy$, compatible with a given
$Y\multi\signal$, is achieved if all multiphoton
events with mode distributions $L_1$ to $L_{K-1}$, and the remaining fraction
 $\lk Y\multi\signal - \sum_{i=1}^{K-1} h_{L_i}\signal \rk/h_{L_K}$ 
with mode distribution $L_K$ are transmitted to Bob's side. As a result we have
the upper bound
\begin{align}
  \label{eq:Ymwc}
  Y\mwc\decoy(Y\multi\signal) &= \sum_{i=1}^{K-1} h_{L_i}\decoy
      + \frac{Y\multi\signal - \sum_{i=1}^{K-1} h_{L_i}\signal}{h_{L_K}\signal}
        h_{L_K}\decoy
\end{align}
on the decoy multiphoton yield $Y\multi\decoy$ for a given signal multiphoton
yield $Y\multi\signal$.

\textbf{Step 3 - The new bound on $Y_1\signal$ for given gains $Q\signal$ and
$Q\decoy$:}
With the derived relations between the yields of signal and idler events, we are
now able to calculate a new lower bound on $Y_1\signal$ for given signal and
decoy gains. If the relations were just given by a constant ratio, this would
be in direct analogy to the single-mode case where the ratio of signal and
decoy yields was fixed. This means we could plug the relations into Equation
\eqref{eq:Qdecoy} and solve Equations \eqref{eq:Qsignal} and \eqref{eq:Qdecoy}
for a lower bound on $Y_1\signal$. However, the derived relations,
Equations \eqref{eq:Yowc} and \eqref{eq:Ymwc}, do not have a simple functional
form. Hence an iterative approach is required to determine the new lower
bound on $Y_1\signal$.

We first solve Equations \eqref{eq:Qsignal} and \eqref{eq:Qdecoy} for
$Y\multi\signal$ and $Y_1\decoy$, respectively:
\begin{align}
  \label{eq:Yms}
  Y\multi\signal &= \frac{1}{P\multi\signal} \left[
    Q\signal - P_0\signal Y_0 - P_1\signal Y_1\signal
    \right], \\
  \label{eq:Y1d}
  Y_1\decoy &= \frac{1}{P_1\decoy} \left[
    Q\decoy - P_0\decoy Y_0 - P\multi\decoy Y\multi\decoy
    \right].
\end{align}
Alice and Bob know $Y_0$ from the vacuum decoy state. They can measure
$Q\signal$ and $Q\decoy$, and they know $P_0$, $P_1$ and $P_M$ for both signal
and decoy because
they know the properties of their source. In addition, a trivial lower bound on
$Y_1\signal$ is given by $Y_1\signal \geq Y\olb\signal = 0$. Starting with that
value, a tighter bound can be calculated by the following algorithm:

\begin{enumerate}
  \item Start by calculating an upper bound on $Y\multi\signal$ from
    $Y_{1,\LB}\signal$, using Equation~\eqref{eq:Yms}:
    \begin{equation}
      Y\mub\signal =
        \frac{1}{P\multi\signal} \left[
        Q\signal - P_0\signal Y_0 - P_1\signal Y\olb\signal
        \right]
    \end{equation}
  \item Next, use $Y\mub\signal$ to derive an upper bound on $Y\multi\decoy$
    with Equation \eqref{eq:Ymwc}
    \begin{equation}
      Y\mub\decoy = Y\mwc\decoy\lk Y\mub\signal\rk
    \end{equation}
  \item Obtain a lower bound on $Y_1\decoy$ from $Y\mub\decoy$
    with Equation~\eqref{eq:Y1d}:
    \begin{equation}
      \label{eq:Y1lbdec}
      Y\olb\decoy =
        \frac{1}{P_1\decoy} \left[
        Q\decoy - P_0\decoy Y_0 - P\multi\decoy Y\mub\decoy
        \right]
    \end{equation}
  \item Finally, determine a lower bound on $Y_1\signal$ from $Y\olb\decoy$,
    using Equation \eqref{eq:Yowc}
    \begin{equation}
      \label{eq:Y1lb}
      Y\olb\signal = Y\owc\signal\lk Y\olb\decoy\rk
    \end{equation}
\end{enumerate}
The value obtained in Equation~\eqref{eq:Y1lb} can iteratively be plugged into
the previously described steps as initial value, which results 
in an even tighter bound. After each iteration step, the final
value for $Y\olb\signal$ is at least as large as the starting value, so the
iteratively obtained values are monotonically increasing.
As $Y\olb\signal \leq 1$ is bounded from above (not more than 100\% of
the events can result in a click of Bob's detector), the series converges,
giving the final lower bound on the single-photon yield of the
signal state.

\section{New Bound on $e_1\signal$}
\label{section:e1}
We also need to bound the error rate of the single-photon events of the signal
state from above. In this case, Eve wants to introduce as many errors as
possible into the single-photon events of the signal state while leaving the
measured QBERs as expected, because this way she
can gain the maximal amount of information from the signal single-photon events.
An upper bound on the decoy single-photon events is given by
Equation~\eqref{eq:e1ub} if we use $Y\olb\decoy$
given by the value \eqref{eq:Y1lbdec} obtained in the iteration for
determining $Y\olb\signal$. Since the errors also have to be assumed to be
under Eve's control, she is free to choose the modes into which the errors
occur. The highest error rate of the signal single-photon events compared to
the error rate of the decoy single-photon events is obtained if the errors are
introduced into modes with a large $m_k\signal/m_k\decoy$ ratio. Hence, 
if we again define K implicitly by
\begin{equation}
  \sum_{\mathclap{k=K}}^N m_k\decoy \leq e_{1,\ub}\decoy
    < \sum_{i=K-1}^N m_k\decoy,
\end{equation}
the worst case assumption is that all photons in modes
$N, N-1, \ldots, K$
and a fraction $\lk e_{1,\ub}\decoy - \sum_{k=K}^N m_k\decoy
\rk/m_{K-1}\decoy$ of the photons in mode $K-1$ are erroneous. This gives the
new upper bound on the signal single-photon error rate,
\begin{equation}
  \label{eq:e1ubsig}
  e_{1,\ub}\signal = \sum_{\mathclap{k=K}}^N m_k\signal
    + \frac{e_{1,\ub}\decoy - \sum_{k=K}^N m_k\decoy}{m_{K-1}\decoy}
        m_{K-1}\signal.
\end{equation}

\section{Numerical Simulations}
\label{section:simulation}
With the new lower bound on $Y_1\signal$, obtained by Equation \eqref{eq:Y1lb}
in the iteration, and the new upper bound on $e_1\signal$ given by
Equation \eqref{eq:e1ubsig}, we can determine a lower bound on the
achievable key rate for states of the form \eqref{eq:rhosigdec} with
Equation~\eqref{eq:keyrate}.

In the following simulations, we consider the simplest example for the use of a
PDC source in a QKD protocol. Signal and decoy states are
created with different pump intensities, and only the photons in the
signal mode are used for information encoding, while the photons in the idler
mode are ignored. 
The PDC output state is modeled with a full multimode structure
and is given by Equation
\eqref{eq:mmpdcstate}, resulting in the state \eqref{eq:rhosigdec} after
tracing over the idler mode. The photon number distribution for each spectral
mode is given by
\begin{equation}
  \label{eq:psigdec}
    p\sigdec(k,n) = \sech^2 r_k\sigdec \tanh^{2n} r_k\sigdec,
\end{equation}
with $r_k\sigdec \propto \lambda_k \sqrt{I\sigdec}$ describing the
corresponding squeezing parameters. $I\signal$ and $I\decoy$
are the pump intensities for signal and decoy state, respectively, and
$\lambda_k$ indicates how prominent the $k$\tth mode is. The coefficients
$\lambda_k$ are properties of the PDC crystal and the pump.

\begin{figure}[t]
  \centering
  \includegraphics[width = 0.45\textwidth]{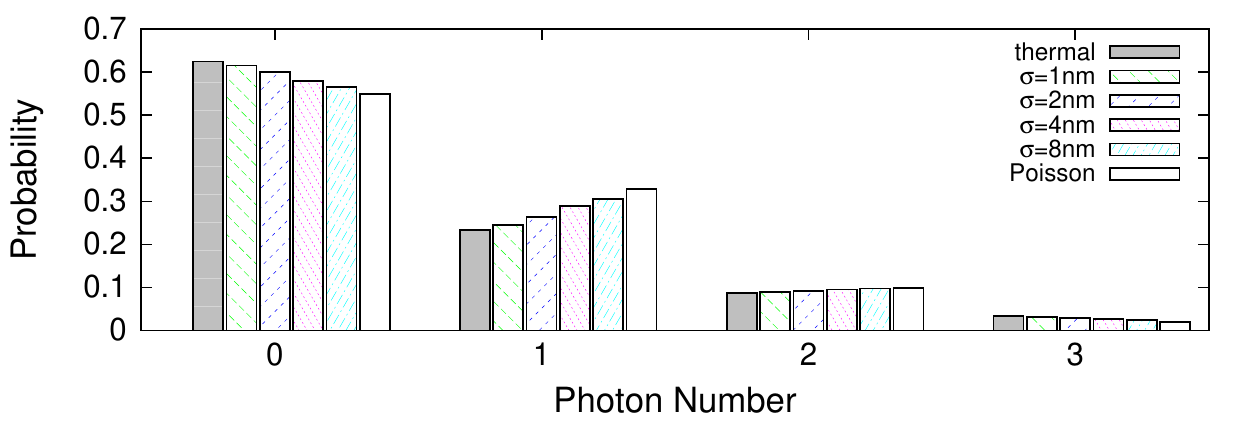}
  \caption{(Color online) The photon number distributions for different pump
widths $\sigma$.}
  \label{fig:statistics}
\end{figure}

The photon number distribution for each spectral mode,
Equation~\eqref{eq:psigdec}, is a thermal distribution. Thus the single-mode
case ($\lambda_k = \delta_{k1}$) corresponds to thermal photon number
distribution. With more
contributing modes, the distribution
is changed from thermal towards a Poissonian distribution. This is shown in
Figure~\ref{fig:statistics} for the values of
Table~\ref{tab:schmidtEV} and a mean photon number of 0.6.

\begin{figure}[tb]
  \centering
  \includegraphics[width = 0.4\textwidth]{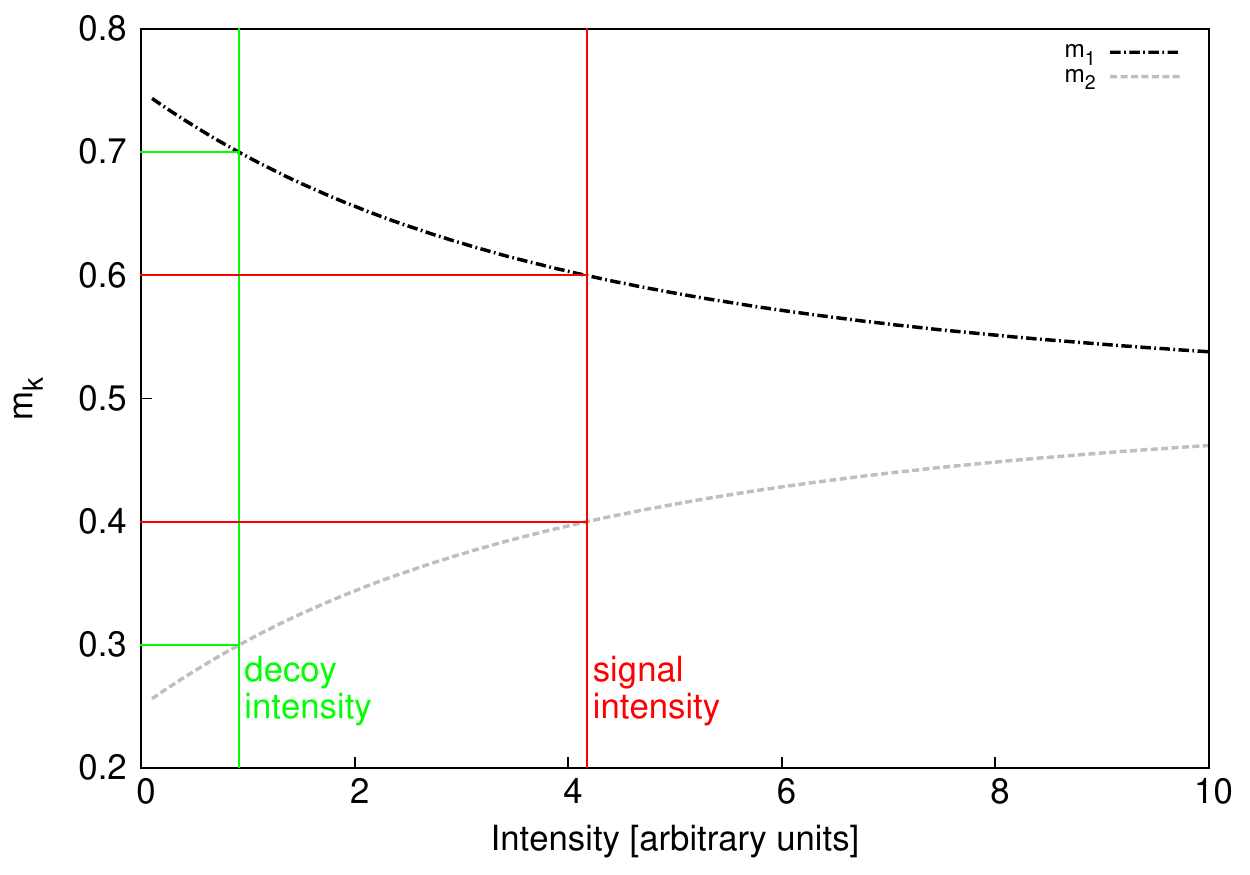}
  \caption{(Color online) The mode occupation probabilities $m_1$ and $m_2$ for
a
    single-photon event in dependence of the pump intensity.}
  \label{fig:mkplots}
\end{figure}

Let us first illustrate, by means of a simple example, how different mode
occupation probabilities arise for signal and decoy state. Consider a
PDC state with just two spectral modes that have $\lambda_1 = \sqrt{0.75}$ and
$\lambda_2 = \sqrt{0.25}$. The mode occupation probabilities for a
single-photon event are then given by
\begin{align}
  m_1(I) &= \frac{p(1,1) p(2,0)}{p(1,1) p(2,0) + p(1,0) p(2,1)}\notag\\
    &= \frac{\tanh^2 r_1}{\tanh^2 r_1 + \tanh^2 r_2},\\
  m_2(I) &= 
     \frac{\tanh^2 r_2}{\tanh^2 r_1 + \tanh^2 r_2}.
\end{align}
They inherit the intensity dependence from the squeezing parameters $r_1$ and
$r_2$. This intensity
dependence is shown in Figure~\ref{fig:mkplots}, along with chosen decoy
and signal intensities such that we end up exactly with the states shown in
Figure~\ref{fig:blocksignalmode}.

Now, we focus on a physically realistic case. Our source is a waveguided
periodically poled KTP crystal with a grating period of $\Lambda = 68.40
\mu$m, length of 5 mm, and waveguide width and height both 4 $\mu$m.  The pump
laser spectrum is centered at a wavelength of 775nm and the signal and idler
are frequency degenerate around 1550nm.  We study four different pump
bandwidths $\sigma$, which lead to different values for $\lambda_k$
\cite{Mauerer2008}. They are shown in Table~\ref{tab:schmidtEV}.

\begin{table}
  \centering
  \begin{tabular}{l|l}
    Width $\sigma$ & $\lambda_k$\\\hline
    1 nm & 0.959, 0.194, 0.152, 0.098, 0.088, 0.033, 0.032, 0.014
        \\\hline
    2 nm & 0.871, 0.463, 0.140, 0.064, 0.054, 0.028, 0.001
        \\\hline
    4 nm & 0.690, 0.555, 0.383, 0.222, 0.107, 0.054, 0.050, 0.044, 
        \\
         & 0.023, 0.012, 0.004, 0.003, 0.001
        \\\hline
    8 nm & 0.511, 0.478, 0.427, 0.364, 0.296, 0.228, 0.167, 0.117, 
        \\
         & 0.078, 0.056, 0.047, 0.037, 0.023, 0.015, 0.014, 0.011, 
        \\
         & 0.006, 0.003, 0.001
  \end{tabular}
  \caption{$\lambda_k$ for the KTP crystal for different
    pump widths.}
  \label{tab:schmidtEV}
\end{table}

We first consider the case of $\sigma=4$nm. For given pump
intensities for signal and decoy state, the mode occupation probabilities can
be calculated with Equations \eqref{eq:mksigdec} and \eqref{eq:hl} for the
single-photon and multiphoton state, respectively. We assume that Eve designs
her attack such that her presence cannot be detected. This implies
that the
measured gains and error rates for signal and decoy state have the values that
are expected from natural losses and detection errors. 
According to \cite{Lo2005}, they are given by
\begin{equation}
  \label{eq:QKD:gain}
  Q\sigdec = \sum_n Y_n P_n\sigdec,
\end{equation}
\begin{equation}
  \label{eq:QKD:QBER}
  E\sigdec = 
    \frac{1}{Q\sigdec} \sum_n e_n Y_n P_n\sigdec,
\end{equation}
with the overall (i.e., convoluted) photon number distributions $P_n\sigdec$
and
\begin{equation}
  \label{eq:Yn}
  Y_n \approx \eta_n + p_\textrm{dark},
\end{equation}
\begin{equation}
  \label{eq:en}
  e_n = \frac{1}{Y_n}
      \lk e_\textrm{det} \eta_n + \frac{1}{2} p_\textrm{dark} \rk.
\end{equation}
In these equations, $p_\textrm{dark}$ is the dark count probability of Bob's
detector, $e_\textrm{det}$ is the detection error (i.e., the probability that
Alice prepares a 0 (1), but Bob detects a 1 (0)),
and $\eta_n = 1-
(1-\eta)^n$ is the probability that at least one of $n$ photons arrives at
Bob's side and is detected. The overall detection probability $\eta =
10^{-\alpha / 10} \cdot \eta_{\mathrm{det}}$ of each
photon is determined by the channel attenuation $\alpha$ in dB and the detector
efficiency $\eta_{\mathrm{det}}$. We use the experimental parameters of
Ref.~\cite{Gobby2004} in the simulations, which are shown in
Table~\ref{tab:expdata}.
\begin{table}[htb]
\centering
\begin{tabular}{l|c}
  Dark count probability \hspace*{0.6cm}& 
      \hspace*{0.3cm}$1.7\cdot10^{-6}$\\\hline
  Detection error & 3.3\%\\\hline
  Detector efficiency & 4.5\%
\end{tabular}
\caption{Characteristics of Bob's detector.}
\label{tab:expdata}
\end{table}

\begin{figure}[tb!]
  \centering
  \includegraphics[width = 0.47\textwidth]{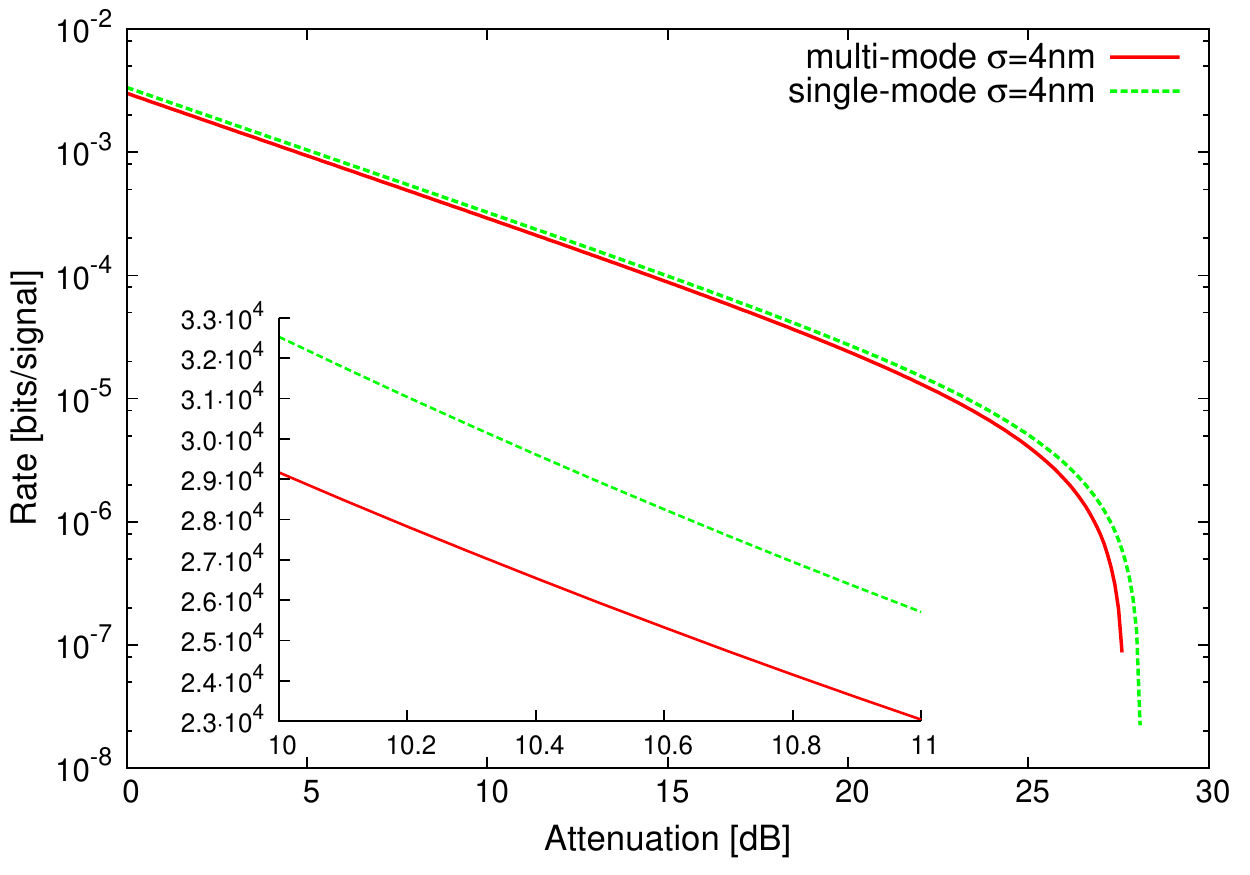}
  \caption{(Color online) Solid (red) line: Lower bound on the secure key rate
    for the KTP crystal with a pump width of 4 nm. 
    Dashed (green) line: Lower bound on secure key rate for the same photon
    number
    distribution, but with all
    photons in the same spectral mode. The inset shows a zoom of the 50-55km
    region. The weak decoy mean photon number is 0.1 in both cases, and
    the signal mean photon number is optimized to result in the largest
    possible rate.}
  \label{fig:mm-vw}
\end{figure}

With the gains and QBER for signal and decoy states, Equations \eqref{eq:Yn}
and \eqref{eq:en}, a lower bound on the signal single-photon yield
$Y_1\signal$, and an upper bound on the signal single-photon error rate
$e_1\signal$ can be calculated as described in Sections \ref{section:Y1} and
\ref{section:e1}. This allows us to compute a lower bound on the achievable key
rate according
to Equation~\eqref{eq:keyrate}. We compare this key rate to the key rate for a
single-mode source with the same photon number
distribution. 
In other words, the secure key rate one would falsely expect to be achievable 
if the multi-mode structure of the PDC state is ignored.
It is calculated by Equation \eqref{eq:keyrate} with the bounds
on $Y_1\signal$ and $e_1\signal$ given by Equations~\eqref{eq:Y1sm} and
\eqref{eq:e1ub}.
The corresponding key rates are both plotted in Figure~\ref{fig:mm-vw} against
the channel attenuation. We find that the key rate drops
about 10\% when Eve's new possible attack is taken into account by adjusting
the bounds on $Y_1\signal$ and $e_1\signal$ accordingly. In both scenarios, the
mean photon number of the decoy state is 0.1, and the mean photon number of the
signal state is optimized to give the highest key rate.

\begin{figure}[tb!]
  \centering
  \includegraphics[width = 0.47\textwidth]{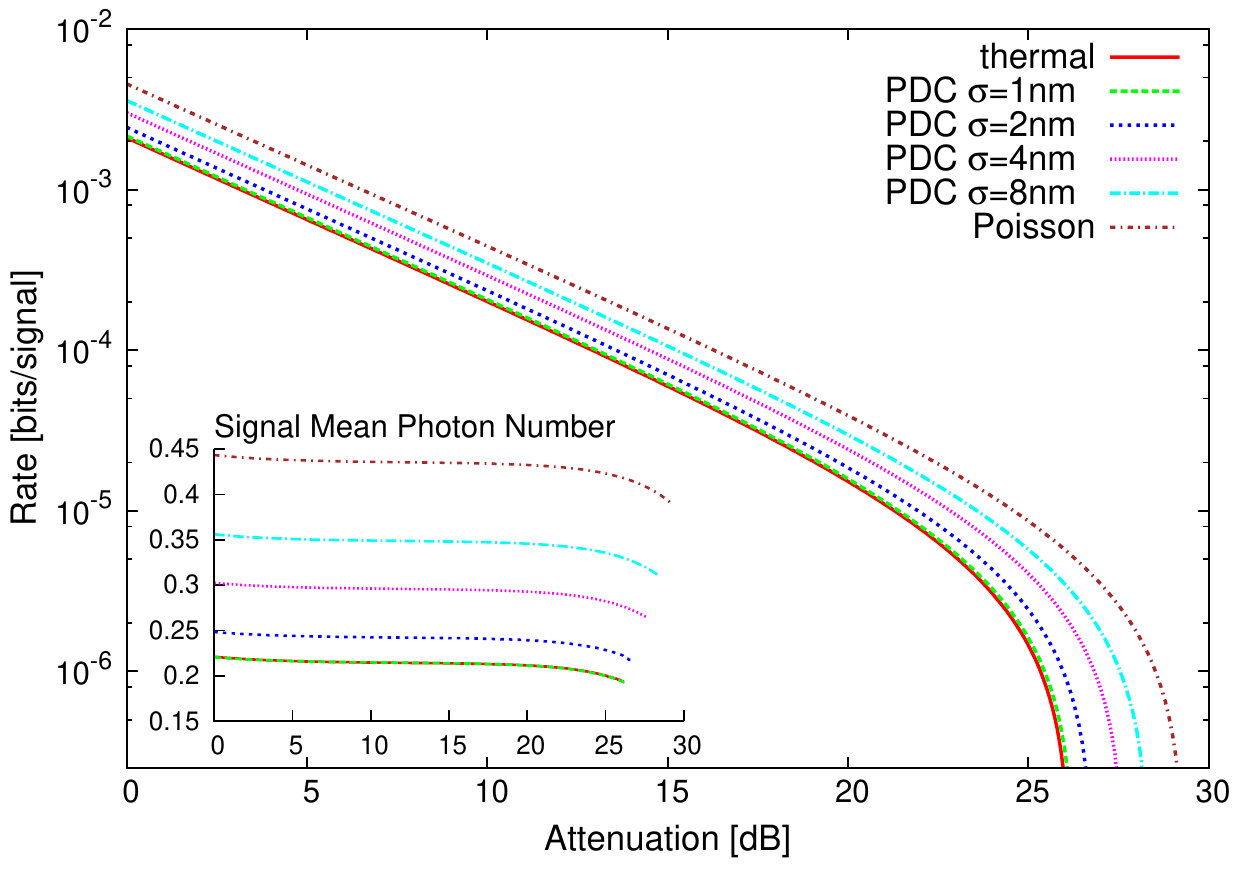}
  \caption{(Color online) Lower bound on the secure key rate for different pump
    width and therefore different mode contributions. The asymptotic cases
    are the rates
    for thermal and Poissonian distribution. The inset shows the optimized mean
    photon number of the signal state. The weak decoy mean photon number is
    fixed to 0.1 for all scenarios.}
  \label{fig:diffsigma}
\end{figure}

Figure~\ref{fig:diffsigma} shows the secure key rate
for all different pump widths given
in Table~\ref{tab:schmidtEV}. One can see that the secure key rate is higher
when more modes contribute to the PDC process. This effect is
explained by the change in the photon number distribution. With more
contributing modes, the photon number distribution is shifted from a thermal
distribution to a Poissonian distribution (see Figure~\ref{fig:statistics}).
The Poissonian distribution is favorable in comparison to the thermal
distribution, because the ratio between
single-photon and multiphoton events increases. This permits a higher
mean photon number for the signal state, which in turn increases the achievable
key rate and distance. The resulting optimal mean photon numbers in dependence
of the channel attenuation are depicted in the inset of
Figure~\ref{fig:diffsigma}.

\section{Conclusion}
\label{sec:conclusion}
In summary, we have pointed out the necessity to carefully pay attention
to the output states of the utilized sources, but likewise demonstrated that the
demand for perfect indistinguishability of the signal and decoy photons, which
is hard to implement in practice, can be loosened for only a small cost in the
key rate.

The analysis was applied to a parametric down conversion
(PDC) source, where the weak decoy state is created by pumping the crystal with
a lower pump intensity. For about ten effectively contributing modes, we
observed
a drop of the key rate to roughly 90\% of the corresponding value in the
single-mode case with the same photon number distribution.
The simulation was performed for different numbers of
effectively contributing modes, leading to the conclusion that the
advantageous change in the photon number distribution, which occurs if more
modes contribute has a higher effect
on the key rate than the aforementioned decrease due to the new attack
possibility presented to Eve by the multi-mode structure of the states.

This analysis can also be used for a heralded PDC source, as long as
the heralding detector is frequency independent, as the resulting states are
also of the form of
Equation~\eqref{eq:rhosigdec}.
For heralding with a frequency dependent
detector, the analysis has to be extended to the case where the
density matrices for signal and decoy state are diagonal in different bases,
contrary to our condition given by Equation~\eqref{eq:rhosigdec}.

Another possibility to produce the decoy state is by passive decoy generation
\cite{Mauerer2007, Adachi2007}. In this scheme, the complications that arise
because of the
multi-mode structure of the PDC state can be avoided if a frequency independent
detector is available for the decoy generation, as the spectral properties of
$n$-photon states would then be the same for signal and decoy state.
This, however, is not the case for a frequency dependent detector, because
such a detector leads to signal and decoy states with different
spectral properties. Again, the resulting states require an analysis for signal
and decoy states that are diagonal in different bases.

We believe that this paper is a first step towards allowing more general signal
and decoy states, which will significantly simplify the design of QKD sources.

\section*{Acknowledgments}
This work was supported by the EC under the FET-Open grant agreement
CORNER, number FP7-ICT-213681

\bibliography{bibtex}

\begin{thebibliography}{10}

\bibitem{Bennett1984}
C.~H. Bennett and G. Brassard,  in {\em Proc. IEEE Int.~Conf.~on Computers,
  Systems, and Signal Processing} (IEEE, New York, ADDRESS, 1984), pp.\
  175--179.

\bibitem{Gisin2002}
N. Gisin, G. Ribordy, W. Tittel, and H. Zbinden, Rev. Mod. Phys. {\bf 74},  145
   (2002).

\bibitem{Duvsek2006}
M. Du\v{s}ek, N. L{\"u}tkenhaus, and M. Hendrych, Progress in Optics {\bf 49},
  381  (2006).

\bibitem{Mauerer2008a}
W. Mauerer, W. Helwig, and C. Silberhorn, Ann. Phys. (Leipzig) 17, No. 2-3 {\bf
  158},  175  (2008).

\bibitem{Mayers1998}
D. Mayers, Journal of the ACM {\bf 3},  35  (1998).

\bibitem{Shor2000}
P.~W. Shor and J. Preskill, Phys. Rev. Lett. {\bf 85},  441  (2000).

\bibitem{Gottesman2004}
D. Gottesman, H.-K. Lo, N. L\"utkenhaus, and J. Preskill, Quant.~Inf.~Comp.
  {\bf 5},  325  (2004).

\bibitem{Lutkenhaus2000}
N. L\"utkenhaus, Phys. Rev. A {\bf 61},  052304  (2000).

\bibitem{Brassard2000}
G. Brassard, N. L\"utkenhaus, T. Mor, and B.~C. Sanders, Phys. Rev. Lett. {\bf
  85},  1330  (2000).

\bibitem{Lutkenhaus2002}
N. L{\"u}tkenhaus and M. Jahma, New J. Phys. {\bf 4},  44.1  (2002).

\bibitem{Hwang2003}
W.-Y. Hwang, Phys. Rev. Lett. {\bf 91},  057901  (2003).

\bibitem{Lo2005}
H.-K. Lo, X. Ma, and K. Chen, Phys. Rev. Lett. {\bf 94},  230504  (2005).

\bibitem{Wang2005a}
X.-B. Wang, Phys. Rev. Lett. {\bf 94},  230503  (2005).

\bibitem{Ma2005}
X. Ma, B. Qi, Y. Zhao, and H.-K. Lo, Phys. Rev. A {\bf 72},  012326  (2005).

\bibitem{Mauerer2008}
W. Mauerer, M. Avenhaus, W. Helwig, and C. Silberhorn, arXiv:0812.3597v1
  (2008).

\bibitem{Wootters1982}
W.~K. Wootters and W.~H. Zurek, Nature {\bf 299},  802  (1982).

\bibitem{Mosley2008}
P.~J. Mosley {\it et~al.}, Physical Review Letters {\bf 100},  133601  (2008).

\bibitem{Rohde2007}
P.~P. Rohde, W. Mauerer, and C. Silberhorn, New Journal of Physics {\bf 9},  91
   (2007).

\bibitem{Gobby2004}
D. Gobby, Z. Yuan, and A. Shields, Appl. Phys. Lett. {\bf 84},  19  (2004).

\bibitem{Mauerer2007}
W. Mauerer and C. Silberhorn, Phys. Rev. A {\bf 75},  050305 (R)  (2007).

\bibitem{Adachi2007}
Y. Adachi, T. Yamamoto, M. Koashi, and N. Imoto, Phys. Rev. Lett. {\bf 99},
  180503  (2007).

\end{thebibliography}
  
\end{document}